\documentclass[10pt,notitlepage,pra,twocolumn,showpacs]{revtex4}%
\usepackage{amsfonts}
\usepackage{amsmath}
\usepackage{amssymb}
\usepackage{graphicx}
\usepackage[toc,page,header]{appendix}%
\providecommand{\U}[1]{\protect\rule{.1in}{.1in}}

\begin{document}
\title{Phase diagram of Rydberg atoms in a nonequilibrium optical lattice}
\author{Jing Qian, Guangjiong Dong, Lu Zhou and Weiping Zhang}
\affiliation{Quantum Institute for Light and Atoms, Department of Physics, East China
Normal University, Shanghai 200062, People's Republic of China}

\begin{abstract}
We study the quantum nonequilibrium dynamics of ultracold three-level atoms
trapped in an optical lattice, which are excited to their Rydberg states via a
two-photon excitation with nonnegligible spontaneous emission. Rich quantum
phases including uniform phase, antiferromagnetic phase and oscillatory phase
are identified. We map out the phase diagram and find these phases can be
controlled by adjusting the ratio of intensity of the pump light to the
control light, and that of two-photon detuning to the Rydberg interaction
strength. When the two-photon detuning is blue-shifted and the latter ratio is
less than 1, bistability exists among the phases. Actually, this ratio
controls the Rydberg-blockade and antiblockade effect, thus the phase
transition in this system can be considered as a possible approach to study
both effects.

\end{abstract}

\maketitle

\textit{Introduction: }Rydberg atoms with principal quantum number $n\gg1$
have exaggerated atomic properties including strong dipole-dipole interactions
and long radiative lifetimes \cite{Gallagher94}. These properties are
attractive for quantum information processing \cite{Lukin01,Saffman10} and
quantum many-body dynamics simulation \cite{Buluta09,Weimer10}. Most of these
researches require a negligible spontaneous emission for reducing the quantum
decoherence \cite{Pupillo10}. However, a recent research shows that when
spontaneous emission is significant, quantum nonequilibrium dynamics could be
demonstrated using Rydberg atom gases \cite{Tony11}. The system investigated
is like a spin-$1/2$ particle system, which undergoes a phase transition from
a spatially uniform phase to an antiferromagnetic phase by tuning the laser
frequency as shown in Ref. \cite{Tony11}. Moreover, the nonequilibrium induced
by the spontaneous emission leads to an oscillatory phase. Further research
has found collective quantum jump between a state of low Rydberg population
and a state of high Rydberg population \cite{Lee12}. These researches open up
a new window to use Rydberg atoms for quantum nonequilibrium dynamics
simulation \cite{Diehl08}.

When two atoms are close, the excitation of one atom is prohibited by an
already excited neighboring atom due to the level shift by the strong
dipole-dipole interaction at a short distance \cite{Viteau11}. This phenomenon
is called dipole blockade effect and has played an important role in current
Rydberg atom researches \cite{Gallagher08,Urban09,Daniel10}. However, recent
theoretical and experimental investigations have shown that in a three-level
two-photon Rydberg excitation scheme, there also exists antiblockade
effect\ due to the Aulter-Townes splitting induced by the lower transition
\cite{Ates07,Amthor10}. The possibility of enhancing the antiblockade effect
by trapping atoms within a lattice has been proposed \cite{Amthor10}. So far,
the exploration of physical effect of the coexistence of blockade effect and
antiblockade effect on quantum dynamics of Rydberg atom gases in this two-step
excitation scheme still remains at its early stage.

In this paper, we propose to study the quantum nonequilibrium dynamics of
three-level atoms trapped within a lattice via a two-step excitation. Adopting
the mean field approach used in Ref. \cite{Tony11}, we predict transition
between uniform phase, antiferromagnetic phase, and oscillatory phase.
Compared to the two-level system investigated in Ref. \cite{Tony11}, we have
more control knobs, including the ratio of intensity of the pump light to the
control light and that of two-photon detuning to the Rydberg interaction
strength, to control the phases. When the latter ratio is less than 1 and the
two-photon detuning is blue-shifted, bistability exists among the phases. The
latter ratio controls the competition of the blockade with antiblockade
effect. Thus, both effects could be investigated through the phase transitions.%

\begin{figure}[ptb]%
\centering
\includegraphics[
height=1.5497in,
width=2.7423in
]%
{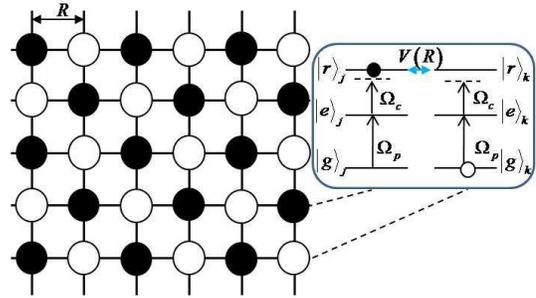}%
\caption{Atoms trapped in an optical lattice are uniformly excited to the
Rydberg state $\left\vert r\right\rangle $ from the ground state $\left\vert
g\right\rangle $ via an intermediate state $\left\vert e\right\rangle $. The
lattice is divided into two sublattices, marked respectively in white and
black colors. The neighboring atoms are coupled through the Rydberg
interaction $V$. $\Omega_{p}$ and $\Omega_{c}$ are two Rabi frequencies
respectively corresponding to two optical transitions.}%
\label{lattice}%
\end{figure}

\textit{Model:} Our scheme is shown in Fig. \ref{lattice}. Ultracold atoms are
trapped within an optical lattice. We assume the lattice depth is deep enough,
so that the center-of-mass motion of the atoms can be neglected. We also
assume there is exactly one atom per site. All the atoms are uniformly excited
by two laser beams to a Rydberg state. The laser-atom coupling scheme is shown
in the right side of Fig. \ref{lattice}, where $\left\vert g\right\rangle $,
$\left\vert e\right\rangle $ and $\left\vert r\right\rangle $ denote the
ground, intermediate and Rydberg states, respectively, for the $j$-th atom,
and $\Omega_{p\left(  c\right)  }$ is the Rabi frequency for the pump
(control) laser. For convenience, we divide the lattice into two sublattices:
the black sublattice and white sublattice, as shown in Fig. \ref{lattice}. We
label the black/white site with $j/k$.

Similar configuration is used in Ref. \cite{Tony11}, where the intermediate
state $\left\vert e\right\rangle $ is adiabatically eliminated for a large
detuning of the $\left\vert g\right\rangle \rightarrow\left\vert
e\right\rangle $ transition from pump laser frequency, and an effective
two-level model can be adopted \cite{quantum}. In contrast, in our paper, a
resonant pump laser is assumed to couple the ground state $\left\vert
g\right\rangle $ and the intermediate state $\left\vert e\right\rangle $. Such
a three-level model allows both the Rydberg-blockade effect and the
antiblockade effect to be studied simultaneously. Whereas, in the effective
two-level model, only the Rydberg-blockade was studied \cite{Urban09}.

In the interaction picture, the Hamiltonian $H_{T}$ describing such a lattice
of atoms can be given by ($\hbar=1$) $H_{T}=\sum_{j}H_{j}+V\sum_{\left\langle
jk\right\rangle }\left\vert r\right\rangle \left\langle r\right\vert
_{j}\otimes\left\vert r\right\rangle \left\langle r\right\vert _{k}$. Here,
$H_{j}$ describing the interaction between the $j$-th three-level atom and two
lasers is given by $H_{j}=-\delta\left\vert r\right\rangle \left\langle
r\right\vert _{j}+(\Omega_{p}\left\vert g\right\rangle \left\langle
e\right\vert _{j}+\Omega_{c}\left\vert e\right\rangle \left\langle
r\right\vert _{j}+h.c.)$ with $\delta$ the two-photon detuning between states
$\left\vert g\right\rangle _{j}$ and $\left\vert r\right\rangle _{j}$. $V$ is
the dipole-dipole interaction between the two Rydberg atoms described by a van
der Waals (vdW) potential.

The quantum dynamics for this system is governed by the master equation of its
density operator $\mathbf{\rho}$ \cite{Gardiner91}:%

\begin{equation}
\mathbf{\Dot{\rho}}=-i\left[  H_{T},\mathbf{\rho}\right]  +\gamma\sum
_{j}(\left\vert g\right\rangle \left\langle e\right\vert _{j}\mathbf{\rho
}\left\vert e\right\rangle \left\langle g\right\vert _{j}-\frac{\{\left\vert
e\right\rangle \left\langle e\right\vert _{j},\mathbf{\rho\}}}{2}),
\label{density_matrix}%
\end{equation}
Here, decoherence due to the spontaneous emission of the intermediate state
$\left\vert e\right\rangle _{j}$ with its linewidth $\gamma$ is
phenomenologically introduced. Dissipation from the spontaneous emission of
Rydberg state $\left\vert r\right\rangle _{j}$ is ignored due to its
relatively long lifetime ($\sim100\mu$s) compared to $\gamma^{-1}$
\cite{Saffman10}.

Now we first investigate the steady-state solutions for Eq.
(\ref{density_matrix}) using the mean field approximation \cite{MF}. With this
approximation, for the $j$-th atom, the Rydberg interaction $\left\vert
r\right\rangle \left\langle r\right\vert _{j}\otimes\sum_{k}\left\vert
r\right\rangle \left\langle r\right\vert _{k}$ ($k$ sums over all the nearest
neighboring sites of the $j$-th site) can be replaced by a formula of density
operator $\left\vert r\right\rangle \left\langle r\right\vert _{j}\sum_{k}%
\rho_{k,rr}$, thus the motional equations for the density operator elements
$\rho_{j,xy}$ ($x,y=g,e,r$) are
\begin{subequations}
\label{pq_sim_eqs}%
\begin{align}
\Dot{\rho}_{j,rr}  &  =2\Omega_{c}\operatorname{Im}\rho_{j,er}\\
\Dot{\rho}_{j,ee}  &  =2\Omega_{c}\left(  \chi\operatorname{Im}\rho
_{j,ge}-\operatorname{Im}\rho_{j,er}\right)  -\rho_{j,ee}\\
\dot{\rho}_{j,ge}  &  =i\Omega_{c}\left(  \chi\left(  1-\rho_{j,rr}%
-2\rho_{j,ee}\right)  +\rho_{j,gr}\right)  -\frac{\rho_{j,ge}}{2}\\
\dot{\rho}_{j,er}  &  =i\Omega_{c}(\rho_{j,ee}-\rho_{j,rr}-\chi\rho
_{j,gr})-i\delta_{eff}\rho_{j,er}-\frac{\rho_{j,er}}{2}\\
\dot{\rho}_{j,gr}  &  =i\Omega_{c}\left(  \rho_{j,ge}-\chi\rho_{j,er}\right)
-i\delta_{eff}\rho_{j,gr}%
\end{align}

In deducing Eqs.(\ref{pq_sim_eqs}), we have scaled the frequencies and time
respectively with $\gamma$ and $\gamma^{-1}$. We have also introduced the
ratio of the pump Rabi frequency to the control Rabi frequency, $\chi
\equiv\Omega_{p}/\Omega_{c}$, and the effective two photon detuning
$\delta_{eff}\equiv\delta-V\sum_{k}\rho_{k,rr}$. Without the Rydberg-Rydberg
interaction $V$, Eqs. (\ref{pq_sim_eqs}) are the typical optical Bloch
equations. The role of the Rydberg-Rydberg interaction here is to compensate
the two-photon detuning $\delta$, as shown in the formula for $\delta_{eff}$,
and thus an antiblockade effect \cite{Ates07,Amthor10} can be induced when
$\delta_{eff}\approx0$.

We now start to investigate the stationary solution of Eqs.(\ref{pq_sim_eqs}).
These stationary solutions can be put into three classes:

(a) Uniform phase (uni): The atoms are uniformly excited over the whole
lattice, i.e., $\rho_{j,rr}^{s}=\rho_{k,rr}^{s}$. Hereafter, the superscript
$s$ stands for the stationary solution.

(b) Antiferromagnetic phase (AF): One sublattice has higher excitation than
the other, i.e., $\rho_{j,rr}\neq\rho_{k,rr}$.

(c) Oscillatory phase (osc): Rydberg population oscillates periodically in
time between two sublattices. In this case, one cannot find any stable
steady-state AF solutions.

We first study the uniform phase ($\rho_{j\left(  k\right)  ,xy}^{s}=\rho
_{xy}^{s})$. From the stationary solution of Eqs.(\ref{pq_sim_eqs}), we obtain
a cubic equation for $\rho_{xy}^{s}$:%
\end{subequations}
\begin{align}
&  \left(  \rho_{rr}^{s}\right)  ^{3}-\frac{2\delta}{V}\left(  \rho_{rr}%
^{s}\right)  ^{2}+\frac{1}{V^{2}}(\delta^{2}+\frac{4\Omega_{c}^{4}\left(
\chi^{2}+1\right)  ^{2}}{1+8\chi^{2}\Omega_{c}^{2}})\rho_{rr}^{s}%
\label{cubic_eq}\\
&  -\frac{4\Omega_{c}^{4}\chi^{2}\left(  \chi^{2}+1\right)  }{V^{2}\left(
1+8\chi^{2}\Omega_{c}^{2}\right)  }=0\nonumber
\end{align}

For AF phase, two sublattices are Rydberg-populated differently, and labeled 1
and 2 in the subscript of $\rho_{j\left(  k\right)  ,xy}^{s}$. From
Eqs.(\ref{pq_sim_eqs}), we obtain equations involving $\rho_{1,rr}^{s}$ and
$\rho_{2,rr}^{s}$:%
\begin{equation}
\rho_{1,rr}^{s}+\frac{\left(  1+8\Omega_{c}^{2}\chi^{2}\right)  \left(
V\rho_{2,rr}^{s}-\delta\right)  ^{2}}{4\Omega_{c}^{4}\left(  1+\chi
^{2}\right)  ^{2}}\rho_{1,rr}^{s}=\frac{\chi^{2}}{1+\chi^{2}}, \label{5}%
\end{equation}%
\begin{equation}
\rho_{2,rr}^{s}+\frac{\left(  1+8\Omega_{c}^{2}\chi^{2}\right)  \left(
V\rho_{1,rr}^{s}-\delta\right)  ^{2}}{4\Omega_{c}^{4}\left(  1+\chi
^{2}\right)  ^{2}}\rho_{2,rr}^{s}=\frac{\chi^{2}}{1+\chi^{2}}. \label{6}%
\end{equation}
The solutions $\rho_{1\left(  2\right)  ,rr}^{s}$ of the coupled equations
(\ref{5}) and (\ref{6}) correspond to two roots of the following equation%
\begin{equation}
\left(  \rho_{rr}^{s}\right)  ^{2}-2A\rho_{rr}^{s}+B=0, \label{quadratic_eq}%
\end{equation}
where $A=\delta/V+2\Omega_{c}^{4}\chi^{2}\left(  1+\chi^{2}\right)  /\left[
4\Omega_{c}^{4}\left(  1+\chi^{2}\right)  ^{2}+\delta^{2}\left(  1+8\chi
^{2}\Omega_{c}^{2}\right)  \right]  $, $B=\delta^{2}/V^{2}+4\Omega_{c}%
^{4}\left(  1+\chi^{2}\right)  ^{2}/\left[  V^{2}\left(  1+8\chi^{2}\Omega
_{c}^{2}\right)  \right]  $.

When $A^{2}-B<0$, the quadratic equation (\ref{quadratic_eq}) has complex
roots. In this situation, the AF and oscillatory phases don't exist, and only
uniform phase exists. In other cases, the quadratic equation
(\ref{quadratic_eq}) has one pair of real solutions. When the real solutions
are stable against small perturbation, the system is in the AF phase;
otherwise, the system is in oscillatory phase.

\begin{figure}[ptb]%
\centering
\includegraphics[
height=1.3136in,
width=3.3831in
]%
{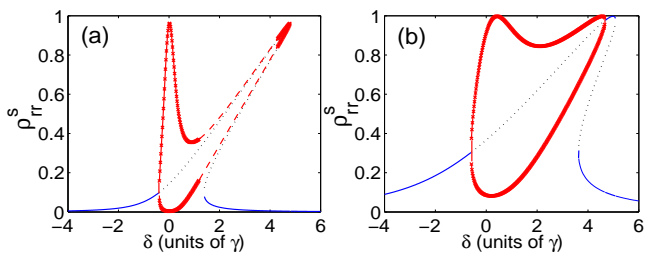}%
\caption{(color online) Stationary solutions $\rho_{rr}^{s}$ as a function of
detuning $\delta$ with $V=5$, $\Omega_{c}=0.1$ for different ratios (a)
$\chi=5$ and (b) $\chi=20$. The stable and unstable uniform phases
respectively show in solid line and dotted line. AF phase and oscillatory
phase are repsectively shown in solid line with cross symbols, and dashed
line.}%
\label{phase _tot}%
\end{figure}

\textit{Phase Diagram}: In this section, we show the numerical results for the
stationary solution with the experimentally tunable two-photon detuning
$\delta$ for $\chi=5$ (Fig. \ref{phase _tot}a) and $20$ (Fig. \ref{phase _tot}%
b). Stable and unstable uniform phases governed by Eq. (\ref{cubic_eq}) are
respectively plotted in solid lines and dotted lines. AF phases and
oscillatory phases from Eq. (\ref{quadratic_eq}) are respectively plotted in
solid lines with cross symbols and dashed lines.

In Fig. \ref{phase _tot}(a), there exists only one stable uniform phase for
small detunings $\delta\leq\delta_{c}\approx-0.5$,. The critical point
$\delta_{c}$ can be determined by $A^{2}=B$. When $\delta$ crosses the
critical point, the uniform phase becomes unstable, giving rise to a pair of
AF phases. This is the pitchfork bifurcation of supercritical type
\cite{Rasband90}, in which the uniform fixed point loses its stability and two
stable AF fixed points appear. As $\delta$ increases further to the value
larger than $1.22$, the AF phase becomes unstable and the system enters into
the oscillatory phase.

The phases can be controlled experimentally by adjusting the ratio of the pump
Rabi frequency to the control one, $\chi$. With a very strong pump transition
($\Omega_{p}\gg\Omega_{c}$), the Rydberg-excitation rate can not catch up with
the decay rate from intermediate state, so that population in Rydberg state
hardly changes. In Fig. \ref{phase _tot}(b) with $\chi=20$\ much larger than
that in Fig. \ref{phase _tot}(a), the stable AF fixed points always exist and
there is no oscillatory phase. There is a supercritical pitchfork bifurcation
at $\delta=4.67$ where the uniform phase becomes stable again.

Figures \ref{phase _tot}(a) and (b) also show the signals of Rydberg-blockade
effect and antiblockade effect. When $\delta=0$, $\left\vert \rho_{1,rr}%
^{s}-\rho_{2,rr}^{s}\right\vert \approx1$ for AF phase. This means a resonant
two-photon detuning leads to an entire Rydberg excitation in one sublattice,
whereas the other sublattice is strongly blocked. When $\delta$ increases to
$5.0$, $\rho_{1,rr}^{s}=\rho_{2,rr}^{s}\approx1.0$,\ and the two-photon
detuning $\delta$ is compensated by the Rydberg energy shift ($\delta
_{eff}\approx0$), indicating the antiblockade effect.%

\begin{figure}[ptb]%
\centering
\includegraphics[
height=1.5584in,
width=2.6066in
]%
{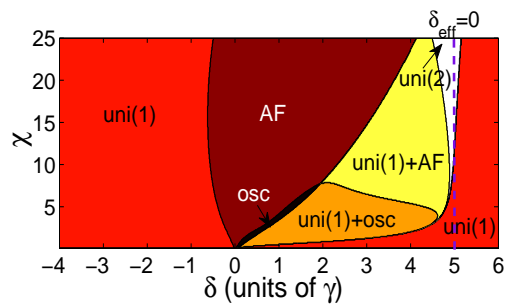}%
\caption{(color online)Phase diagram within ($\delta,\chi$) space for $V=5$,
$\Omega_{c}=0.1$. }%
\label{Phase_DT}%
\end{figure}

Since the stationary solutions are dependent on both the two-photon detuning
$\delta$ and the ratio of the Rabi frequencies for pump and control lasers
$\chi$, we plot the phase diagram in the parameter space of $(\delta$, $\chi)$
in Fig. \ref{Phase_DT}. The number of stable uniform phases is denoted in a
bracket. The position that $\delta_{eff}=0$ has been marked. When
$\delta_{eff}\gtrsim0$ ($\delta\gtrsim V\sum_{k}\rho_{k,rr}$), the Rydberg
interaction plays negligible role, and thus the system behaves like a
noninteracting particle system. Consequently, there is only one stable uniform
phase, no matter how $\chi$ varies. When $\delta$ is comparable to $V\sum
_{k}\rho_{k,rr}$, there is complex phase distribution within $(\delta$,
$\chi)$ space. AF phase contains a pair of stable nonuniform fixed points. In
the regions marked with "uni(2)", "uni(1)+AF" and "uni(1)+osc", bistability
takes place between two uniform phases, or uniform and antiferromagnetic phase
or uniform and oscillatory phase. When the two-photon detuning $\delta$ is
resonant ($\delta=0$), the system is in AF phase. When $\delta$ is red shifted
($\delta<0$) and near-resonant ($\delta\approx0$), there could be AF phase or
uniform phase depending on the ratio $\chi$; however, if $\delta$ is further
red-shifted, there is only uniform phase.%

\begin{figure}[ptb]%
\centering
\includegraphics[
height=1.4909in,
width=2.5036in
]%
{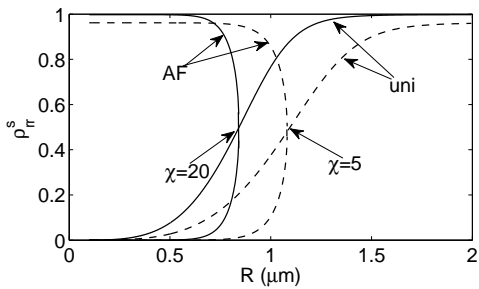}%
\caption{(color online)Stationary solutions $\rho_{rr}^{s}$ as a function of
lattice spacing $R$ for $\chi=5$ (dashed lines) and $\chi=20$ (solid lines). }%
\label{R_phase}%
\end{figure}

\textit{Rydberg-Rydberg interaction}: The phase diagram analysis has shown
that the Rydberg-Rydberg interaction strength could have strong influence on
the dynamics of Rydberg gases. In experiments, the Rydberg-Rydberg interaction
strength can be controlled with the atom-atom separation $R$. Now we take an
isotropic vdW interaction $V=-dC_{6}/\gamma R^{6}$($d$ is the lattice
dimension) for short separation $R$ as an example to investigate the
controlling of the Rydberg gas dynamics with the lattice size
\cite{Osychenko11,Ji11}. For $^{87}$Rb atoms,\ we choose $5s$, $5p$ and
Rydberg state respectively as $\left\vert g\right\rangle _{j}$, $\left\vert
e\right\rangle _{j}$ and $\left\vert r\right\rangle _{j}$. The interaction
coefficient for Rydberg atoms is $C_{6}=-870$kHz$\mu$m$^{6}$ when $\left\vert
r\right\rangle _{j}=23s_{1/2}$ \cite{Reinhard07}, and linewidth $\gamma$ from
state $\left\vert e\right\rangle _{j}$ is $\gamma/2\pi=1.7$ MHz
\cite{Pritchard10}. Using these parameters, the normalized vdW potential is
$V\approx1/R^{6}$ ($R$ is in units of $\mu$m, $d=2$). In this case, we obtain
stationary solutions $\rho_{rr}^{s}$ with $\delta=0$ as a function of spacing
$R$ as shown in Fig. \ref{R_phase} for $\chi=5$ (dashed lines) and $20$ (solid lines).%

When $R\ll1.0\mu$m, the system is dominated by a pair of AF phase with one
sublattice completely Rydberg-excited ($\rho_{rr}^{s}\approx1$) and the other
blockade($\rho_{rr}^{s}\approx0$). This is a demonstration of the Rydberg
blockade phenomenon under strong Rydberg interaction \cite{Gaetan09}. From
experimental view, $R\ll1.0\mu$m can be realized in solid systems, but it is
hard for atoms trapped within a conventional optical lattice \cite{Kubler09}.
However, subwavelength optical lattices can be facilitated with a recent
progress in subwavelength focusing, and have been realized recently for matter
wave Bloch oscillation \cite{Salger09}. Thus, observing AF phase is promising
in lattices created with subwavelength focusing techniques.

When $R$ increases, the uniform phase develops and one sees the two-phase (uni
and AF) coexistence in the regime around $R\sim1\mu$m. After crossing over
this regime, the AF phase disappears and the uniform phase dominates with $R$
increasing further, due to the negligible value of $V$ for large atom-atom separations.%

\begin{figure}[ptb]%
\centering
\includegraphics[
height=2.1525in,
width=3.5198in
]%
{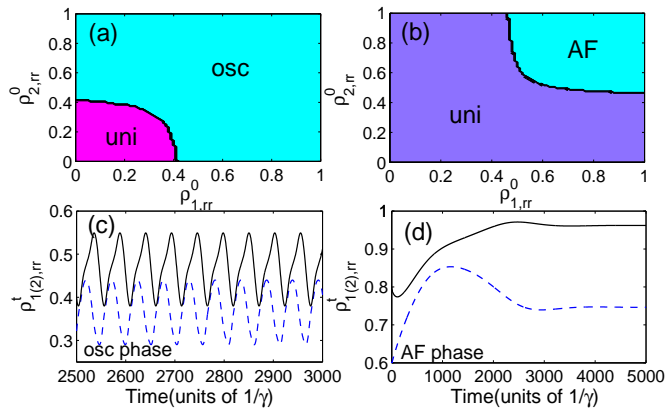}%
\caption{(color online) Phase diagram in the space of initial conditions
$\left(  \rho_{1,rr}^{0},\rho_{2,rr}^{0}\right)  $ with (a) $\chi=5$,
$\delta=2$ and (b) $\chi=20$, $\delta=4$. (c) and (d) show typical Rydberg
population dynamics of oscillatory phase ($\rho_{1,rr}^{0}=0.4$, $\rho
_{2,rr}^{0}=0.6$) and AF phase ($\rho_{1,rr}^{0}=0.6$, $\rho_{2,rr}^{0}=0.8$),
respectively.}%
\label{phase_diagram_ini}%
\end{figure}

\textit{Dynamics}: The dynamical evolution of the system is explored via a
straightforward time-integration of the mean-field equations (\ref{pq_sim_eqs}%
). According to the phase diagram of Fig. \ref{Phase_DT}, the system is in the
bistable state between the uniform phase and oscillatory phase for $\chi=5$,
$\delta=2$.\ The results of dynamical simulation in Fig.
\ref{phase_diagram_ini}(a) gives the information about which phase the system
will finally evolve into by choosing different initial Rydberg populations
($\rho_{1,rr}^{0},\rho_{2,rr}^{0}$). If both $\rho_{1,rr}^{0}$ and
$\rho_{2,rr}^{0}$ are smaller than $0.4$, the final state remains in the
uniform phase. By increasing $\rho_{1,rr}^{0}$ or $\rho_{2,rr}^{0}$, due to
the bistability effect, the system enters into oscillatory phase with Rydberg
population periodically oscillating between two sublattices(see (c) for
$\rho_{1,rr}^{0}=0.4$, $\rho_{2,rr}^{0}=0.6$). Fig. \ref{phase_diagram_ini}(b)
is obtained in the same way except with $\chi=20$, $\delta=4$. Under this set
of parameters, the system is in the bistable state between uniform phase and
AF phase according to Fig. \ref{Phase_DT}. The uniform phase now becomes
dominated. This is because with a far-off resonant two-photon detuning and a
weak control laser, the transition from intermediate state to Rydberg state
becomes very difficult, leading to small Rydberg population and in turn weak
Rydberg interaction. Therefore the system tends to stay in the uniform phase.
However, as long as $\rho_{1,rr}^{0}$ and $\rho_{2,rr}^{0}$ are both
significantly large, the system will still evolve into the AF phase. Fig.
\ref{phase_diagram_ini}(d) presents a typical Rydberg population dynamics in
the case of AF phase for $\rho_{1,rr}^{0}=0.6$, $\rho_{2,rr}^{0}=0.8$.

\textit{Conclusions}: We have studied mean-field phase diagrams in a lattice
system of three-level atoms with two-photon Rydberg excitation and spontaneous
emission-induced dissipation. We have shown that the system has rich phases,
including antiferromagnetic phase, uniform phase, and oscillatory phase. We
can control the phases using either the ratio of the intensity of the pump to
the control light, or that of two-photon detuning to the Rydberg interaction
strength. The latter ratio actually tunes the competition of the
Rydberg-blockade effect and the antiblockade effect. We found when the Rydberg
interaction strength $V$ is much larger than two-photon detuning $\delta$ (the
Rydberg-blockade dominates), the system can be in the AF phase. When $V$ is
comparable to $\delta$, the competition of the Rydberg-blockade eand the
antiblockade effects leads to bistability of the system. When $\left\vert
\delta\right\vert $ is far larger than $V$, the system is in the uniform
phase. Our work shows the possibility of studying antiblockade effect within a
lattice \cite{Amthor10}.

We thank T. Lee for useful discussions. This work was supported by National
Basic Research Program of China (973 Program) under Grant No. 2011CB921604,
2011CB921602, the NSFC under Grant Nos. 11104076, 11004057, 11034002, 10974057
and 10874045, the Specialized Research Fund for the Doctoral Program of Higher
Education No. 20110076120004, the "Chen Guang" project supported by Shanghai
Municipal Education Commission and Shanghai Education Development Foundation
under Grant No. 10CG24


\begin{thebibliography}{99}                                                                                               %


\bibitem {Gallagher94}T. F. Gallagher, \textit{Rydberg Atoms} (Cambridge
University Press, Cambridge, 1994)..

\bibitem {Lukin01}M. D. Lukin \textit{et al.}, Phys. Rev. Letts. \textbf{87}
037901 (2001).

\bibitem {Saffman10}M. Saffman, T. G. Walker and K. M\O lmer, Rev. Mod. Phys.
\textbf{82} 2313 (2010) and references therein.

\bibitem {Buluta09}I. Buluta and F. Nori, Science \textbf{326} 108 (2009).

\bibitem {Weimer10}H. Weimer \textit{et al.}, Nature Phys. \textbf{6} 382 (2010).

\bibitem {Pupillo10}G. Pupillo \textit{et al.}, Phys. Rev. Letts. \textbf{104}
223002 (2010).

\bibitem {Tony11}T. E. Lee, H. H\"{a}ffner and M. C. Cross, Phys. Rev. A
\textbf{84} 031402 (2011); I. Lesanovsky Physics \textbf{4} 71 (2011).

\bibitem {Lee12}T. E. Lee, H. H\"{a}ffner and M. C. Cross, Phys. Rev. Letts.
\textbf{108} 023602 (2012).

\bibitem {Diehl08}S. Diehl \textit{et al.}, Nature Phys. \textbf{4} 878 (2008).

\bibitem {Viteau11}M Viteau \textit{et al.}, Phys. Rev. Letts. \textbf{107}
060402 (2011).

\bibitem {Gallagher08}T. F. Gallagher and P. Pillet, Advances in Atomic,
Molecular, and Optical Physics \textbf{56} 161 (2008).

\bibitem {Urban09}E. Urban \textit{et al.}, Nature Phys. \textbf{5} 110
(2009); M. Weidem\"{u}ller Nature Phys. \textbf{5} 91 (2009).

\bibitem {Daniel10}D. Comparat and P. Pillet, J. Opt. Soc. Am. B \textbf{27}
A208 (2010), and references therein.

\bibitem {Ates07}C. Ates, T. Pohl, T. Pattard and J. M. Rost, Phys. Rev.
Letts. \textbf{98} 023002 (2007).

\bibitem {Amthor10}T. Amthor, C. Giese, C. S. Hofmann and M. Weidem\"{u}ler,
Phys. Rev. Letts. \textbf{104} 013001(2010).

\bibitem {quantum}Heping Zeng, Weiping Zhang and Fucheng Lin, Phys. Rev. A
\textbf{52} 2155 (1995).

\bibitem {Gardiner91}C. W. Gardiner and P. Zoller, \textit{Quantum Noise}
(Springer-Verlag 2004).

\bibitem {MF}P. M. Chaikin and T. C. Lubensky \textit{Principles of condensed
matter physics} (Cambridge University Press, 2007).

\bibitem {Rasband90}S. N. Rasband, \textit{Chaotic Dynamics of Nonlinear
Systems} (New York: Wiley, 1990)..

\bibitem {Osychenko11}O. N. Osychenko \textit{et al.}, Phys. Rev. A
\textbf{84} 063621 (2011).

\bibitem {Ji11}S. Ji, C. Ates, and I. Lesanovsky, Phys. Rev. Letts.
\textbf{107} 060406 (2011).

\bibitem {Reinhard07}A. Reinhard, T. C. Liebisch, B. Knuffman and G. Raithel,
Phys. Rev. A \textbf{75} 032712 (2007).

\bibitem {Pritchard10}J. D. Pritchard \textit{et al.}, Phys. Rev. Letts.
\textbf{105} 193603 (2010).

\bibitem {Gaetan09}A. Ga\"{e}tan \textit{et.al.} Nat. Phys. \textbf{5} 115 (2009).

\bibitem {Kubler09}H. K\"{u}bler \textit{et al}., Nature Photon. \textbf{4} 112(2010).

\bibitem {Salger09}T. Salger \textit{et al.}, Phys. Rev. A \textbf{79}
011605(R) (2009).
\end{thebibliography}
\end{document}